\documentstyle[aps,pra,twocolumn]{revtex}

\begin{document}
\draft
\title{Temperature-induced resonances and Landau damping of collective
modes in Bose-Einstein condensed gases in spherical traps}
\author{M. Guilleumas$^1$, L.P. Pitaevskii$^{1,2}$}
\address{$^1$Dipartimento di Fisica, Universit\`a di Trento and
Istituto Nazionale per la Fisica della Materia, I-38050 Povo, Italy}
\address{$^2$Kapitza Institute for Physical Problems, 117454 Moscow,
Russian Federation}
\date{June 14, 1999}

\maketitle

\begin{abstract}

Interaction between
collective monopole oscillations  of
a trapped Bose-Einstein condensate and thermal
excitations
is investigated  by
means of perturbation
theory. We assume spherical symmetry to calculate the matrix elements
by solving the linearized Gross-Pitaevskii equations.
We use them to study the
resonances of the condensate induced by temperature when an external
perturbation of the trapping frequency is applied and to calculate the
Landau damping of the oscillations.
\end{abstract}

\section{Introduction}

Since the discovery of Bose-Einstein condensation in magnetically trapped
Bose gases, the study of the low-energy collective excitations has attracted
a big interest both from experimental and theoretical point of view.
Mean field theory has proven to be a good framework to study static, dynamic
and thermodynamic properties of these trapped gases. In particular, it
provides predictions of the frequencies of collective excitations that
very well agree with the observed ones. Recently
\cite{JILA,MIT} the
energy shifts and damping rates of these low-lying collective excitations
have been measured as a function of temperature.
However, these phenomena have not yet been
completely understood theoretically.

In this paper we study
the influence of thermal excitations on collective oscillations of the
condensate in the collisionless regime.
Previous papers on this subject has been devoted mainly to
calculation of the Landau damping by means of perturbation theory.
In Refs.\cite{Liu1,PLA,Liu2,Giorgini1} only
the
uniform system has been considered,
whereas in Refs.\cite{Fedichev1,Fedichev2} Landau damping in trapped
Bose gases has also been studied but using the semiclassical
approximation for thermal excitations and the hydrodynamic
approximation for collective oscillations.
An important point of Refs.\cite{Fedichev1,Fedichev2}
is that the authors discuss the possible
chaotic behavior of the excitations in an anisotropic trap.
The frequency shift has also been studied
for a trapped condensate in the collisionless regime in
Ref.\cite{Fedichev2}.

In the present work we study the interaction between collective and
thermal excitations using the Gross-Pitaevskii equation and perturbation
theory. We consider spherically symmetric traps, since in this case the
spectrum of excitations is easily calculated, avoiding the use of
further approximations. Even though the case of anisotropic traps can be
significantly different in the final results, a detailed investigation
of spherical traps is instructive. We explore, in particular, the
properties of monopole oscillations by studying the temperature-induced
resonances
that occur in the condensate when an external perturbation of the
trapping frequency is applied and, also,
the Landau damping associated with the interaction with
thermally excited states.

This paper is organized as follows. In section II we introduce the
general
equations that describe the elementary excitations of the condensate
within the Bogoliubov theory \cite{bog}. In Sec. III we recall
the perturbation theory for a trapped Bose-condensed gas in order to
study the interaction between
elementary excitations.
In Sec. IV we introduce
the linear response function formalism and calculate the response function
of the condensate when a small perturbation of the trapping frequency is
applied. We derive analytic equations for the response function at zero
temperature and treat perturbatively the  contribution of the elementary
excitations, which is related to Landau damping. In Sec. V we
discuss the main results.

\section{Elementary excitations of an isotropic trap}

We consider a weakly interacting Bose-condensed gas confined in an
external potential $V_{\rm ext}$ at $T=0$.
The elementary excitations of a degenerate Bose gas are  associated
with the fluctuations of the condensate. At low temperature they are
described by the time  dependent Gross-Pitaevskii (GP) equation for the
order parameter \cite{G,P}:
\begin{equation}
 i\hbar {\partial \over \partial t} \Psi ({\bf r},t) =
\left( - { \hbar^2 \nabla^2 \over 2m} + V_{\rm ext}({\bf r})
+ g \mid \!\Psi({\bf r},t) \!\mid^2 \right) \Psi({\bf r},t)  \; ,
\label{TDGP}
\end{equation}
where $\int \! d{\bf r} |\Psi|^2= N_0$ is the number of atoms in
the condensate. At zero
temperature it coincides with the total number of atoms $N$, except
for a very small difference $\delta N \ll N$ due to the
quantum depletion of the condensate.  The coupling constant
$g$ is proportional to the  $s$-wave scattering  length $a$ through
$g=4\pi \hbar^2 a/m$. In the present work we will  discuss the case
of positive scattering length, as for $^{87}$Rb atoms.
The trap is included through $V_{\rm ext}$, which is chosen here in the
form of an isotropic harmonic potential:  $V_{\rm ext}(r)= (1/2) m
\omega_{\rm ho}^2 r^2$. The harmonic trap provides a typical
length scale for the system,
$a_{\rm ho}= (\hbar/m\omega_{\rm ho})^{1/2}$.

So far experimental  traps have axial
symmetry, with different radial and axial frequencies, but experiments
with spherical traps are also feasible \cite{KetVa}. The choice here
of a spherical trap has two different reasons. First, it
greatly reduces the numerical effort
and will allow us to study the
interaction of oscillations with elementary excitations without any
further approximations. Second,
the energy spectrum of the
excitations in such a trap is well resolved yielding
to the appearance of well-separated resonances. In anisotropic traps,
conversely, the spectrum of excitations is much denser.

The normal modes of the condensate can be found by linearizing
equation (\ref{TDGP}) , i.e., looking for solutions of the form
\begin{equation}
\Psi({\bf r},t) = e^{-i\mu t/\hbar} \left[ \Psi_0
({\bf r}) + u({\bf r}) e^{-i \omega t} + v^*({\bf r}) e^{i \omega t}
\right]
\label{linearized}
\end{equation}
where $\mu$ is the chemical potential and functions $u$ and $v$ are
the ``particle" and ``hole" components characterizing the Bogoliubov
transformations. After inserting in Eq.~(\ref{TDGP}) and retaining
terms up to first
order in $u$ and $v$, one finds three equations.
The first one is the nonlinear equation for the
order parameter of the ground state,
\begin{equation}
\left( H_0 + g \Psi_0^2 ({\bf r})  \right) \Psi_0({\bf r})
= \mu   \Psi_0({\bf r}) \, ,
\label{groundstate}
\end{equation}
where $H_0= - (\hbar^2/2m) \nabla^2 +  V_{\rm ext}({\bf r})$;  while
$u({\bf r})$  and $v({\bf r})$ obey the following
coupled equations~\cite{P}:
\begin{eqnarray}
  \hbar \omega u({\bf r}) &=& [ H_0 - \mu + 2 g \Psi_0^2]  u ({\bf r})
+ g  \Psi_0^2 v ({\bf r})
\label{coupled1}
\\
- \hbar \omega v({\bf r}) &=& [ H_0 - \mu + 2 g \Psi_0^2]  v ({\bf r})
+ g  \Psi_0^2 u ({\bf r}) \; .
\label{coupled2}
\end{eqnarray}
Numerical solutions of these equations have been found by
different authors \cite{Edwards,Singh,Esry,Zaremba,You,PRA}. In the
present work, we use them to calculate
the response function of the condensate under an external
perturbation and
the Landau damping of collective
modes.

When the adimensional parameter $N a/a_{\rm ho}$ is large, the
time-dependent GP equation reduces to the hydrodynamic equations
\cite{Stringari}:
\begin{equation}
\frac{\partial \rho}{\partial t}+\nabla({\bf v}\rho)=0
\label{hydron}
\end{equation}
\begin{equation}
m\frac{\partial}{\partial t}{\bf
v}+\nabla(V_{\rm ext}+g\rho+\frac{mv^2}{2})=0\,, \label{hydrov}
\end{equation}
where $\rho({\bf r},t)=\mid \Psi({\bf r},t) \mid^2$
is the particle density and
the velocity field is
${\bf v}({\bf r},t)=(\Psi^* \nabla \Psi-\Psi \nabla \Psi^*)\hbar/(2mi\rho)$.
The static solution of equations (\ref{hydron})-(\ref{hydrov}) gives the
Thomas-Fermi ground state density, which in the spherical symmetric trap reads
\begin{equation}
\rho(r)=g^{-1} [\mu-V_{\rm ext}(r)]
\label{TF}
\end{equation}
in the region where $\mu>V_{\rm ext}(r)$, and $\rho=0$ elsewhere.
The chemical potential $\mu$ is fixed by the normalization of the density
to the number of particles $N_0$ in the condensate.
The density profile (\ref{TF}) has the
form of an inverted parabola, which vanishes at the classical turning
point $R$ defined by the condition $\mu=V_{\rm ext}(R)$. For a spherical
trap, this implies
\begin{equation}
\mu=\frac{m \omega_{\rm ho}^2 R^2}{2} \,.
\label{TFR}
\end{equation}
It has been shown \cite{Stringari} that the hydrodynamic equations
(\ref{hydron}) and (\ref{hydrov}) correctly reproduce the low-lying
normal modes of the trapped gas in the linear regime
when $N a/a_{\rm ho}$ is large
(see however Ref.\cite{hydro}).

\section{Perturbation theory}

Let us briefly recall the perturbation theory
for the interaction between collective modes of a condensate
and thermal excitations
as it was developed in Ref.\cite{PLA}.
Suppose that a certain mode of the condensate has been excited
and, therefore, it oscillates with the corresponding frequency
$\Omega_{\rm osc}$.
We assume
that this oscillation is classical, i.e. the number of quanta of
oscillation ($n_{\rm osc}$) is very large.
Then, the energy of the system associated with the occurrence of this
classical oscillation can be calculated
as $E=\hbar\Omega_{\rm osc}\, n_{\rm osc}$ with $n_{\rm osc}\gg1$.
Due to interaction effects, the thermal bath can either absorb or emit
quanta of this mode producing a damping of the collective oscillation.
The energy loss can be written as
\begin{equation}
\dot{E}=-\hbar \Omega_{\rm osc} (W^{(a)}-W^{(e)}) \,,
\label{eq1}
\end{equation}
where $W^{(a)}$ and $W^{(e)}$ are the probabilities of absorption and
emission of one quantum $\hbar \Omega_{\rm osc}$, respectively.
The interaction between excitations
is small, so one can use perturbation theory to calculate the probabilities
for the transition between a $i$-th excitation and a $k$-th one,
available by thermal activation
\begin{equation}
W=\pi \sum_{i,k} \mid \langle k \mid V_{\rm int} \mid i \rangle \mid^2 \,.
\label{prob}
\end{equation}
Let $E_{i}$ and $E_{k}$ be the corresponding energies
and assume $E_{k}> E_{i}$.
Since energy is conserved during the transition process, one has
$E_{k}=E_{i}+\hbar\Omega_{\rm osc}$.

The interaction term in second quantization is given by
\begin{equation}
V_{\rm int}=\frac{g}{2} \int d{\bf r}\, \hat{\Psi}^{\dag}
\hat{\Psi}^{\dag} {\hat \Psi}{\hat \Psi}  \,.
\label{Vint}
\end{equation}
In the framework of Bogoliubov theory,
the field operator $\hat \Psi$ can be written as the sum of the condensate
wave function $\Psi_0$, which is the order parameter at equilibrium, and
its fluctuations $\delta \hat{\Psi}$, where
$\hat{\Psi}=\Psi_0+\delta \hat{\Psi}$
[see Eq.~(\ref{linearized})].
The fluctuations can be expressed in terms of the annihilation ($\alpha$)
and creation ($\alpha^{\dag}$) operators of the elementary excitations
of the system:
\begin{equation}
\delta\hat{\Psi}=\sum_{j}[u_j({\bf r})\alpha_j +
   v_j^*({\bf r}) \alpha_j^{\dag}] \,,
\label{uv}
\end{equation}
where the functions $u$ and $v$ are
properly normalized solutions of equations
(\ref{coupled1})-(\ref{coupled2}).
In the sum (\ref{uv}) one can select a low energy collective mode, for
which we use the notation
$u_{\rm osc},v_{\rm osc},\alpha_{\rm osc}, \alpha^{\dag}_{\rm osc}$,
and investigate its interaction with higher energy single-particle
excitations, for which we use the indices
$i,k$ as in (\ref{prob}). These latter excitations are assumed to be
thermally excited.
Inserting expression (\ref{uv}) into Eq.~(\ref{Vint}) one rewrites the
interaction term $V_{\rm int}$ in
terms of the annihilation and creation operators.
Since
we want to study the decay process in which a quantum of oscillation
$\hbar \Omega_{\rm osc}$ is annihilated (created) and the $i$-th
excitation is transformed into the $k$-th one (or viceversa),
we will keep only terms linear in
$\alpha_{\rm osc}\,(\alpha^{\dag}_{\rm osc})$ and in
the product $\alpha_k^{\dag} \alpha_i \,(\alpha_k \alpha_i^{\dag})$.
And the energy conservation during the transition process will be ensured
by the delta function $\delta(E_k-E_i-\hbar\Omega_{\rm osc})$.
This mechanism is known as Landau damping
\cite{Beliaev}.

Assuming that at equilibrium the states $i,k$ are thermally occupied
with the usual Bose factor $f_i=[\exp(E_i/k_{B}T)-1]^{-1}$,
the rate of energy loss can be calculated as \cite{PLA}
\begin{equation}
\dot{E}=-2 \pi \frac{E}{\hbar} \sum_{ik} \mid A_{ik}\mid^2
  \delta(E_k-E_i-\hbar\Omega_{\rm osc})(f_i-f_k) \,,
\label{dE/dt}
\end{equation}
where
\begin{eqnarray}
A_{ik}&=&2g\int d{\bf r}\, \psi_{0}[(u_k^* v_i+v_k^* v_i+u_k^* u_i)
u_{\rm osc}  \nonumber \\
      & & + (v_k^* u_i+v_k^* v_i+u_k^* u_i)v_{\rm osc}].
\label{matrixel}
\end{eqnarray}

Let us define the dissipation rate $\gamma$ through the following relation
between the energy of the system $E$ and its dissipation $\dot{E}$:
\begin{equation}
\dot{E}=-2\gamma E \,.
\label{dampdef}
\end{equation}
Using expression (\ref{dE/dt}) $\gamma$ can be calculated as
\begin{equation}
\frac{\gamma}{\Omega_{\rm osc}}=\sum_{ik} \gamma_{ik} \,
\delta(\omega_{ik}-\Omega_{\rm osc})\,,
\label{damping2}
\end{equation}
where the transition frequencies
$\omega_{ik}=(E_k-E_i)/\hbar$ are positive. The ``damping strength"
\begin{equation}
\gamma_{ik}=\frac{\pi}{\hbar^2\Omega_{\rm osc}} \mid A_{ik}\mid^2(f_i-f_k)
\label{dampingik}
\end{equation}
has the dimensions of a frequency.
In this work we calculate the quantities $\gamma_{ik}$ by using the
numerical solutions $u$ and $v$ of Eqs.~(\ref{linearized}-\ref{coupled2})
into the integrals (\ref{matrixel}). The results will be discussed
in section V.

\section{response function}

The results of the previous section can be used also to
study the effect that an external perturbation of the trap
has on the collective excitations of the condensate.
Let us assume the trapping frequency in the form
[$\omega_{\rm ho} +\delta \omega_{\rm ho}(t)$], where
$\delta \omega_{\rm ho} \sim \exp(-i \omega t)$ is a
time-dependent modulation.
Assuming that the perturbation is small, one can use the response
function formalism to describe the fluctuations of the system.
Let us briefly recall the basic formalism \cite{Landau}.

The behaviour of a system under an external perturbation can be described
by studying the fluctuations that may generate the external interaction to
a certain physical quantity of the system.
An external perturbation acting on the system is described by a new term in
the Hamiltonian of the type
\begin{equation}
\hat{V}=-\hat{x}f(t)\,,
\label{V}
\end{equation}
where $\hat{x}$ is the
quantum operator of the physical quantity that may fluctuate, and $f(t)$ is
the ``perturbing force". The mean value $\langle x \rangle $
is zero in the equilibrium
state, in absence of perturbation, and is not zero when it is present.
For a periodic perturbation
$f(t) \sim \exp(-i\omega t)$, the relation between $\langle x \rangle $ and
$f(\omega)$ is
\begin{equation}
\langle x \rangle =\alpha(\omega)f \,,
\label{respons}
\end{equation}
where $\alpha(\omega)$ is the response function also called generalised
susceptibility.

In general $\alpha(\omega)$ is a complex function.
It can be seen that the imaginary part of the susceptibility
determines the absorption of energy $Q$ of the external force $f$
by the system through the following relation:
\begin{equation}
Q=\frac{\omega}{2} Im[\alpha(\omega)]\, |f|^2 \,,
\label{dissipation}
\end{equation}
and that the real and imaginary parts of $\alpha(\omega)$ satisfy the
Kramers-Kr\"onig relation
\begin{equation}
Re[\alpha(\omega)]=\frac{2}{\pi} \,P \int_{0}^{\infty}
\frac{Im[\alpha(\xi)]}{\xi^2-\omega^2}\,\xi\, d\xi \,,
\label{KramersKronig}
\end{equation}
where $P$ means the principal value of the integral.

 The time-dependent external drive $\delta \omega_{\rm ho}$
induces oscillations of the condensate
density $\delta \rho $
with frequency $\omega$; $\rho(r,t)=\rho(r,0)+\delta\rho$.
Expanding the energy due to the confining potential,
$E_{\rm ho}=\int V_{\rm ext} \,\rho \, d{\bf r}$, with
respect to  $\delta \omega_{\rm ho}$ and $\delta \rho $ one obtains
the ``mixed''
term, corresponding to the Hamiltonian (\ref{V}):
\begin{equation}
V=m \omega_{\rm ho} \delta\omega_{\rm ho} \int r^2 \delta \rho \,d{\bf r}\,.
\label{V2}
\end{equation}
Comparing it with Eq.~(\ref{V}) one can identify
the perturbing force and the corresponding coordinate as
\begin{equation}
f=-m \omega_{\rm ho} \delta \omega_{\rm ho} \,\,,
\,\,\, x=\int r^2 \delta \rho(r,t) d{\bf r}\, .
\label{f}
\end{equation}
Note that the first order term $m\omega _{\rm ho}\delta \omega_{\rm ho}
\int r^2 \rho(r,0)d{\bf r}$
can be omitted because it gives an additive shift in
the Hamiltonian
which does not contribute to the equations of motion of the system.

Once we have identified $f$ and $x$,
we can calculate the response function of the condensate
$\alpha(\omega)$. According to the definition one has
\begin{equation}
x=\alpha(\omega) f \,.
\label{alphadef}
\end{equation}
Let us present the response function in the form
$\alpha(\omega)=\alpha_0(\omega)+\alpha_1(\omega)$, where
$\alpha_0(\omega)$ corresponds to the response function of the
condensate at $T=0$, i.e., calculated without elementary
excitations, and $\alpha_1(\omega)$ is the contribution of the excitations.
At low temperatures it can be assumed that
$\alpha_1(\omega) \ll \alpha_0(\omega)$ and then $\alpha_1$
can be treated as a perturbation.

We proceed as follows.
First, we use the hydrodynamic approximation
to obtain  the response function at $T=0$.
Then, within a perturbation theory,
we introduce the contribution of the elementary excitations
at finite $T$ to obtain $\alpha_1(\omega)$.

\subsection{Calculation of $\alpha_0(\omega)$ at $T=0$}

For a spherically symmetric breathing mode \cite{Lev},
one can easily prove that the hydrodynamic
equations of motion (\ref{hydron}) and
(\ref{hydrov}) admit analytic solutions of the form \cite{Concetta1}
\begin{equation}
\rho(r,t)=a_0(t)-a_r(t)r^2\,\,,
\,\,\,v(r,t)=\alpha_r(t) r\,.
\label{n(r,t)}
\end{equation}
These equations
are restricted to the region where $\rho\geq 0$.
Notice that they include the ground state
solution (\ref{TF}) in the Thomas-Fermi limit. This is recovered by putting
$\alpha_r=0$, $a_r=m\omega_{\rm ho}^2/(2g)$, and $a_0=\mu/g$.
Inserting Eqs.~(\ref{n(r,t)}) into the hydrodynamic
equations, one obtains two coupled differential equations for the time
dependent coefficients $a_r(t)$ and $\alpha_r(t)$, while at any time
$a_0=-(15 N/8\pi)^{2/5} a_r^{3/5}$ is fixed by the normalization of
the density to the total number of atoms.
The form (\ref{n(r,t)})
for the density and velocity distributions is equivalent to a
scaling transformation of the order parameter. That is,
at each time, the parabolic
shape of the density is preserved, while the classical radius $R$, where the
density (\ref{n(r,t)}) vanishes, scales in time as \cite{Dalfovo}
\begin{equation}
R(t)=R(0)\, b(t)=\sqrt{\frac{2\mu}{m\omega_{\rm ho}^2}} \, b(t)\,,
\label{R(t)}
\end{equation}
where the unperturbed radius $R(0)$ is given by
Eq.~(\ref{TFR}).

The relation between the scaling parameter $b(t)$ and the
coefficient $a_r(t)$ is $a_r=m\omega_{\rm ho}^2/(2gb^5)$. Inserting it into
Eq.~(\ref{n(r,t)}) we obtain
\begin{equation}
\rho(r,t)=-\frac{1}{g}\left[m\omega_{\rm ho}^2 r^2 \frac{1}{2b^5}-
	\mu \frac{1}{b^3} \right] \,.
\label{n(r,t)2}
\end{equation}
And the hydrodynamic equations then yield  $a_r=\dot{b}/b$ and
\begin{equation}
\ddot{b}+[\omega_{\rm ho}+\delta \omega_{\rm ho}(t)]^2 \,b
-\frac{\omega_{\rm ho}^2}{b^4}=0\,.
\label{b(t)}
\end{equation}
The second and third terms of (\ref{b(t)}) give the effect of the external
trap and of the interatomic forces, respectively.
From (\ref{R(t)}) and (\ref{b(t)}) it follows that at
equilibrium $b=1$ and $\dot{b}=0$.
For a small driving strength $\delta \omega_{\rm ho}$, one can assume that
the radius of the cloud is
perturbed around its equilibrium value, so
\begin{equation}
R(t)=R(0)+\delta R(t) \,\,,
\,\,\,b(t)=1+\delta b(t)\,,
\label{dR}
\end{equation}
where
\begin{equation}
\delta b(t)=\frac{\delta R(t)}{R(0)}\,.
\label{fract}
\end{equation}
It means that
$\delta b$ is the fractional amplitude of oscillations of the radius
and, therefore, it is a measurable quantity.

In the small amplitude limit, one can linearize Eq.~(\ref{b(t)}) with
respect to $\delta\omega_{\rm ho}$ and $\delta b$ yielding the
following equation
\begin{equation}
\delta \ddot{b}+5\omega_{\rm ho}^2 \delta b=-2\omega_{\rm ho}
\delta\omega_{\rm ho} \,.
\end{equation}
The solution is
\begin{equation}
\delta b(t)=\frac{-2 \omega_{\rm ho}}{\Omega_{\rm M}^2-\omega^2}\,
\delta\omega_{\rm ho} \,,
\label{db}
\end{equation}
where $\Omega_{\rm M}=\sqrt{5} \, \omega_{\rm ho}$ corresponds to the frequency
of the normal mode of monopole in the hydrodynamic limit
\cite{Stringari}.

Keeping only the lowest
order in the small perturbation $\delta b$,
Eq.~(\ref{n(r,t)2}) yields
\begin{equation}
\rho(r,t)=\rho(r,0)+
  \frac{1}{g}\left[5 \frac{1}{2}m\omega_{\rm ho}^2 r^2-3\mu\right]\delta
b\,, \label{n(r,t)3}
\end{equation}
and using the Thomas-Fermi radius at equilibrium (\ref{TFR}),
it follows that the density fluctuation is given by
\begin{eqnarray}
\delta \rho(r,t)&=&\rho(r,t)-\rho(r,0) \nonumber \\
  & =&\frac{5\mu}{g}\left[\left(\frac{r}{R(0)}\right)^2-\frac{3}{5}\right]
   \delta b(t) \,.
\label{dn}
\end{eqnarray}
We can calculate now $x$ using (\ref{f}) and (\ref{dn}), finding
\begin{equation}
x=C \, \delta b(t) \,,
\label{dx2}
\end{equation}
where $C=16 \pi \mu R(0)^5/(35 g)$. Then, from Eq.~(\ref{db}), one gets
\begin{equation}
x= \frac{-2 \, C \,\omega_{\rm ho}}{\Omega_{\rm M}^2-\omega^2}\,\delta
\omega_{\rm ho} \,.
\label{dx3}
\end{equation}
At $T=0$ there are no thermally excited states and, hence,
$\alpha(\omega)=\alpha_0(\omega)$. By
comparing the definition (\ref{alphadef}) with (\ref{dx3}) one has
\begin{equation}
\alpha_0(\omega)=\frac{2 C}{m(\Omega_{\rm M}^2-\omega^2)}\,.
\label{alpha0}
\end{equation}
This is
the response function at zero temperature
without including any dissipation. Therefore
$\alpha_0(\omega)$ is real, i.e., the induced oscillations at $T=0$
are undamped.

The energy of oscillation can be calculated as twice the mean kinetic
energy associated to the mode, $E=\int d{\bf r}\, \rho(r,0) v^2$.
For a monopole mode in an isotropic trap, the calculation
\cite{Lev,Concetta1} gives $E=\frac{15}{7}N\mu|\delta b|^2$, where
$|\delta b|$ is the amplitude of the oscillation of the cloud
(\ref{fract}). Using Eqs. (\ref{dx2}) and (\ref{alphadef}) at $T=0$, it
follows that
\begin{equation}
E=\frac{15}{7 C^2} \mu N |\alpha_0(\omega) f|^2 \,.
\label{Eosc}
\end{equation}

\subsection{Calculation of $\alpha_1(\omega)$}

Now we want to calculate the contribution of the thermally excited states
to the response function. We study the low temperature regime, where
$\alpha_1 \ll \alpha_0$ and the energy of oscillation (\ref{Eosc}) can be
estimated using $\alpha_0$ instead of $\alpha(\omega)=\alpha_0(\omega)+
\alpha_1(\omega)$. The effect of $\alpha_1$ will be introduced within
a perturbation theory.

We have already seen that the thermal excitations
can either absorb or emit quanta of oscillation $\hbar \omega$ and
thus they will dissipate energy. The contribution of the elementary
excitations to the susceptibility will be a complex function,
$\alpha_1(\omega)= Re[\alpha_1]+{\rm i} Im[\alpha_1]$, whose imaginary part
is related to the absorption of energy $Q$ of the external perturbation.
However, in a stationary solution which is the case
under consideration, the absorption $Q$ must be compensated by
the energy dissipation (\ref{dampdef}) due to the interaction with
the elementary excitations. Therefore,
\begin{equation}
Q+\dot{E}=0 \,.
\label{Q}
\end{equation}
Let us rewrite the definition of the damping rate (\ref{dampdef}) by
using (\ref{damping2}) and (\ref{dampingik}) with a generic oscillation
frequency $\omega$:
\begin{equation}
\dot{E}=-2\omega \sum_{ik} \gamma_{ik} \,
   \delta(\omega_{ik}-\omega) \, E \,.
\label{Edot}
\end{equation}
Inserting Eq.~(\ref{Eosc}) and defining
$\beta(\omega)=\alpha_0(\omega)/C=
   2/[m(\Omega_{\rm M}^2-\omega^2)]$
one obtains the energy dissipation
\begin{equation}
\dot{E}=-2\omega \frac{15 \mu N}{7}\sum_{ik} \gamma_{ik} \,
   \delta(\omega_{ik}-\omega) \, |\beta(\omega)|^2 |f|^2 \,.
\label{Edot2}
\end{equation}
Let us recall
that
the energy
dissipation according to
Eqs. (\ref{dissipation}) and (\ref{Q})
can be calculated also from the imaginary part of the
response
function $\alpha(\omega)=\alpha_0(\omega)+\alpha_1(\omega)$. Since
$\alpha_0(\omega)$ is real, Eq.~(\ref{dissipation}) becomes
\begin{equation}
Q=\frac{\omega}{2} Im[\alpha_1(\omega)]\, |f|^2=-\dot{E} \,.
\label{dissipation2}
\end{equation}
Comparing Eqs.~(\ref{Edot2}) and (\ref{dissipation2}) one can calculate
the imaginary part of $\alpha_1(\omega)$ as
\begin{equation}
Im[\alpha_1(\omega)]= 4 \,\frac{15 \mu N}{7} \sum_{ik}
   \gamma_{ik} \,\delta(\omega_{ik}-\omega) \, |\beta(\omega)|^2\,.
\label{Ima1}
\end{equation}
And using the Kramers-Kr\"onig relation (\ref{KramersKronig}) one finds
the real part
\begin{equation}
Re[\alpha_1(\omega)]=\frac{8}{\pi} \sum_{ik}
    \frac{\omega_{ik} \gamma_{ik}}{\omega_{ik}^2-\omega^2}
    \frac{15 \mu N}{7} |\beta(\omega_{ik})|^2 \,.
\label{Rea1}
\end{equation}
Now we have all the ingredients to calculate the response
function of a spherically symmetric trapped condensate when the
monopole mode is excited and a small
perturbation of the trapping frequency
$\delta \omega_{\rm ho} \sim \exp(-i \omega t)$
is applied. It can be calculated within first order perturbation as
$\alpha(\omega)=\alpha_0(\omega)+Re[\alpha_1(\omega)]
+{\rm i} Im[\alpha_1(\omega)]$, by using Eqs.~(\ref{alpha0}),
(\ref{Rea1}) and (\ref{Ima1}), respectively.
It is worth stressing that
the real part of the susceptibility diverges at $\Omega_{\rm M}$ (resonance
of the condensate at $T=0$) but also at $\omega_{ik}$ which are the
frequencies of the thermal excited modes that due to the interaction
are coupled with the monopole.

Actually, the resonances of the condensate can be found by
measuring the fractional amplitude of oscillations
of the cloud radius $\delta b$ at different perturbing frequencies.
This measurable quantity
can be easily related to the
response function $\alpha(\omega)$ from Eqs.~(\ref{dx2}) and
(\ref{alphadef})
\begin{equation}
\delta b=-\alpha(\omega)\,\frac{m \Omega_{\rm ho}}{C}\,
\delta \omega_{\rm ho}\,.
\label{db2}
\end{equation}

Note that the perturbation theory we have used is valid when
$\mid \alpha
_1\mid \ll \mid \alpha _0\mid$. This condition becomes very restrictive
at $\omega $ near $\omega _M $. However it is not difficult to improve the
approximation in this region
by taking benefit of the analogy
between the response function and the Green function $G$.

It is well known that the Green function obeys the Dyson equation
\cite{Abrikosov}
which relates the perturbed quantity ($G$) and the unperturbed one
($G_0$) through the inversed functions ($G^{-1}$ and $G_0^{-1}$) in such
a way that a perturbation theory for $G^{-1}$ has a more wide
applicability that for $G$.
Analogously, we will find a relation between the inverse response
functions, perturbed ($\alpha^{-1}$) and unperturbed ($\alpha_0^{-1}$).
One has
\begin{equation}
\frac{1}{\alpha}=\frac{1}{(\alpha_0+\alpha_1)}=
 \frac{1}{\alpha_0 (1+\alpha_1/\alpha_0)} \,
\label{alpha-1}
\end{equation}
and formally with the same accuracy
\begin{eqnarray}
\frac{1}{\alpha}&\simeq &\frac{1}{\alpha_0} (1-\frac{\alpha_1}{\alpha_0})
  \nonumber \\
&=&\frac{m}{2 C}(\Omega_{\rm M}^2-\omega ^2)-
\frac{8}{\pi} \sum_{ik}
    \frac{\omega_{ik} \gamma_{ik}}{\omega_{ik}^2-\omega^2}
    \frac{15 \mu N}{7C^2}  \,.
\label{alpha-1b}
\end{eqnarray}
Now the applicability of (\ref{alpha-1b}) is restricted only by the
condition that the second term is small with compare to
$\frac{m}{2 C} \Omega_{\rm M}^2$.

It is worth noting that according to the equation (\ref{alpha-1b}) the poles
of $\alpha (\omega) $ related to the resonances are shifted with compare
to
frequencies $\omega_{ik} $ and are given by the equation:
$\alpha_1(\omega_R')/\alpha_0(\omega_R')=1$. However these
shifts are very small.

\section{RESULTS}

In order to present numerical results we choose a gas of $^{87}$Rb atoms
(scattering length $a=5.82 \cdot 10^{-7}$ cm).
For the spherical trap we fix the frequency $\omega_{\rm ho}=
2\pi 187$ Hz, which is the geometric average
of the axial and radial frequencies
of Ref.~\cite{JILA},
and corresponds to the oscillator length $a_{\rm ho}=0.791 \cdot 10^{-4}$ cm.
We solve the linearized Gross-Pitaevskii equations
Eqs.~(\ref{linearized}-\ref{coupled2}) at zero temperature,
to obtain the
ground state wave function $\Psi_0$ and the
spectrum of excited states $E_i$ as well as the corresponding functions
$u_i({\bf r}),v_i({\bf r})$.
In spherically symmetric traps the eigenfunctions
are labeled by $i=(n,l,m)$, where $n$
is the number of nodes in the radial solution, $l$
is the orbital angular momentum and $m$ its projection.
The eigenfunctions are $u_{nlm}({\bf r})=U_{nl}(r) Y_{lm}(\theta,\psi)$,
the energies $E_{nl}$ are
$(2 l+1)$ degenerate
and the occupation of the thermally excited states is
fixed  by the Bose factor.

For a fixed number of trapped atoms, $N$, the number of atoms in the
condensate, $N_0$, depends on temperature $T$.
At zero temperature all the atoms are in the condensate,
except a negligible quantum depletion \cite{PRA}.
At finite temperature the condensate atoms coexists with the thermal
bath.
In the thermodynamic limit \cite{Giorgini3} the
$T$-dependence of the condensate fraction is $N_0(T)= N[1-(T/T_c^0)^3]$.

We consider the collective excitations in the collisionless regime. This
regime is achieved
at low enough temperature.
The excitation spectrum at low temperature can be safely calculated by
neglecting the coupling between the condensate and thermal
atoms \cite{Popov}.
It means that the excitation
energies at a given $T$ can be obtained within Bogoliubov theory at $T=0$
normalizing the number of condensate atoms to $N_0(T)$.

We investigate the monopole mode
($l=m=0$ and $n=1$). The functions
$u_{\rm osc}$ and $v_{\rm osc}$ do not present angular dependence, and from
Eq.~(\ref{matrixel}) it is straightforward to see that the matrix
element $A_{ik}$ couples only those energy levels ($i,k$) with the same
quantum numbers $l$ and $m$. That is, the selection rules corresponding
to the monopole-like transition are $\Delta l=0$ and $\Delta m=0$.
It is obvious, also, that different pairs of levels with the same
quantum numbers $n$ and $l$
but different $m$ give the same contribution.
Therefore,
only the integration of the radial part has to be done numerically.

Fixed $N_0$ and at a given temperature, we
calculate the damping strengths
(\ref{dampingik}) for the transitions $\omega_{ik}$ coupled with the
monopole. In Figure 1 we show the values of
$\gamma_{ik}$ (in units of the frequency of the monopole $\Omega_{\rm M}$)
for $N_0=50000$ $^{87}$Rb atoms at $k_B T = \mu$.
The arrow points to the frequency of the breathing mode
$\Omega_M=2.231 \,\omega_{\rm ho}$, and the
chemical potential is $\mu=15.69 \,\hbar \omega_{\rm ho}$
[these values are numerical results of
the linearized Gross-Pitaevskii equations
(\ref{linearized})-(\ref{coupled2}) for $N_0=50000$ rubidium atoms].
The position of the bars correspond to the allowed transition frequencies
$\omega_{ik} $ (in units of $\omega_{\rm ho}$)
whereas their height defines the numerical
value of $\gamma_{ik}$ \cite{dipole}.

One can see that there are two
different types of allowed transitions $\omega_{ik}$.
The damping strength associated to most of them is very small.
Conversely, there are a few transitions which give relatively
large values of $\gamma_{ik}$. The latter
correspond to transitions between the lowest levels
($n_k=1,
n_i=0$) for different values of $l$ ($l= 2, 3, 4, 5$).
The main reason for these ``strong transitions'' is that the
temperature occupation factor for these low-lying levels is large.
Moreover the
calculation shows that the matrix elements are also enhanced compared
to other transitions.
This is due to the fact that the radial wave functions involved in the
integration have either one ($n_k=1$) or no node ($n_i=0$),
differently from the oscillating character of the radial
wave functions associated to higher levels \cite{hydro}.

The contribution of the
other transitions is like a small
``background'' which is difficult to resolve in the scale of the Figure.
A close-up view of the damping strengths of the transition frequencies
around the monopole is displayed in the inset of Fig.~1 in order to show
the dense background.
It is worth stressing that such a distinction
between ``background'' and ``strong'' transitions
depends on the  number of condensed atoms in the
system and, of course, on temperature.
When the number of atoms in the condensate increases, the number of
excited states available by thermal excitations also increases, leading
to a denser and less resoluble background.

In Figure 2 we present the same as in Fig. 1 but for $N_0=5 000$
atoms of rubidium at $k_B T = \mu$, where here
$\mu=6.25 \hbar \omega_{\rm ho}$.
In this case, one can see that
the difference between the ``strong'' and ``weak'' transitions is not
so impressive as in a bigger condensate
since all damping strengths can be appreciated in the same scale.

We can conclude that at large $N_0$ we have actually two
different
phenomena. The strong transitions create temperature induced resonances
which can be observed in direct experiments. The background transitions
give rise to Landau damping of the collective oscillations
(See subsection B).

\subsection{Temperature-induced resonances}

Using the transition frequencies $\omega_{ik}$ and the corresponding
damping strength $\gamma_{ik}$, we have calculated the response
function $\alpha(\omega)$.
At zero temperature, the response function $\alpha_0(\omega)$
given by Eq.~(\ref{alpha0}), gives a
resonance at the monopole frequency $\Omega_{\rm M}=\sqrt 5 \omega_{\rm ho}$
evaluated in the hydrodynamic regime.
Due to interaction, thermal excited modes are coupled with the
monopole. It means that when one excites the breathing mode of the
condensate, the elementary excitations can give rise to
other resonances at $\omega_{ik}$, which are the frequencies where
$Re[\alpha_1(\omega)]$ diverges [see Eq.~(\ref{Rea1})].
We will now discuss the conditions for the observation of these effects
in actual experiments. In particular, we calculate
the contribution of these resonances to
the response function and
estimate the associated strengths.

Let us study the resonances at $k_B T= \mu$ for $N_0=150 000$ atoms of
$^{87}$Rb. The behavior of the damping coefficients $\gamma_{ik}$
is analogous to the one for $50 000$ condensate atoms (see Fig.~1)
but in this case the difference between `` strong'' resonances and small
background is even bigger: the dense background is not more resoluble in
the scale of the strong resonances. There are five
resonances that stand out the others, and that we
label as $\omega_{R}$ and $\gamma_{R}$ the corresponding damping
strength
(see table 1 for numerical values).

For perturbing frequencies close to the monopole $\omega \sim \Omega_{\rm M}$,
the monopole susceptibility , Eq.~(\ref{alpha0}), can be
approximated to
\begin{equation}
\alpha_0(\omega)=\frac{2C}{m(\Omega_{\rm M}-\omega)(\Omega_{\rm M}+\omega)}
\simeq A_0\frac{1}{(\Omega_{\rm M}-\omega)}\,,
\label{aproxalpha0}
\end{equation}
where $A_0=C/(m \Omega_{\rm M})$.

Analogously, $\alpha _1(\omega)$ near each resonance
$\omega \sim \omega_{R}$
can be presented in the form $\alpha _1(\omega) \simeq A_{1}/(\omega
_{R}-\omega)$. The
ratio $A_{1}/A_0$
is a measure of the relative intensity between temperature-induced and
monopole resonance.

Table 1 displays the numerical values of the relative intensity
for each temperature-induced resonance $\omega_{R}$ with respect
to the monopole one,
for
$N_0=150 000$ atoms in the condensate at $k_B T= \mu$.
The relative strenght of the response function ($A_{1}/A_0$)
at $\omega_{R}$
depends not only on the damping coefficient $\gamma_{R}$ but also on
$(\Omega_{\rm M}^2-\omega_{R}^2)^{-1}$. It means that one mode
$\omega_{R}$
will be easier to excite, i.e., the strength of the response will be
bigger, when it is close to the frequency of the monopole.
Note also that the resonance strength increases with temperature
through $\gamma_{R}$.

 From table 1 one can see that the biggest resonance occurs
at $\omega_R=2.2576\, \omega_{\rm ho}$ which is resoluble
from the monopole
frequency $\Omega_{\rm M}=2.234\, \omega_{\rm ho}$ and
has a large enough relative strength to be observed.
It means that, tuning the perturbation frequency $\omega$ to this value,
a fluctuation of the fractional
amplitude of oscillations can be observed.

In Figure 3 we have plotted the  frequency dependence of the
real part of response function
$\alpha (\omega)$ calculated according to equation
(\ref{alpha-1b}) for $N_0=150 000$.
The response function is given in arbitrary units, and frequency
is in units of $\omega_{\rm ho}$. The dashed line shows the
monopole resonance at $\Omega_{\rm M}$, whereas the other divergences of
$\alpha(\omega)$ correspond to the temperature-induced resonances at
$\omega_{R}$. From this figure one can see
that the thermal induced resonances are quite distinct one from
the other and from the monopole one.
Therefore, temperature-induced resonances could be observed in
experiments with good enough frequency resolution
and good accuracy in the measurement of the radius fluctuations.

We would like to stress that the phenomenon we have discussed is
related to quite delicate
features of interaction between elementary excitations, and therefore,
its observation
would give rich information about properties of Bose-Einstein condensed
gases at finite temperature.

\subsection{Landau damping of collective modes}

 From Fig. 1 one can see that the weak background
transitions $\omega_{ik}$
have, generally speaking, very small frequency separation.
To estimate this distance quantitatively let us renominate the
resonances by an index $i$ in the order of increasing value of
$\omega$. Then, one can define the average distance between
resonances $\overline{\Delta \omega }$ according to:
\begin{equation}
\overline{\Delta \omega} = \frac{\sum_{i} \gamma _i(\omega
_{i+1}-\omega_i)}{\sum_{i} \gamma _i} \,.
\label{av}
\end{equation}
In a small interval around the collective oscillation
$0.82\, \Omega_{\rm M}< \omega_{ik}<1.18 \,\Omega_{\rm M} $,
we sum up all the transition frequencies allowed by the
monopole selection rules and find the following values for
the average distance between two consecutive transition frequencies:
$\overline{\Delta \omega} /\omega _{\rm ho}\simeq 0.0006, 0.001$ and $0.006$
for $N_0=150 000$, $50 000$ and  $5 000$, respectively.
It is hopeless, of course, to try to
resolve these resonances. Actually, there are reasons to believe that these
resonances are smoothed and overlapped. First of all,
because a real trap cannot be
exactly isotropic. It means that levels with different $m$ have
slightly different energies, only levels with $m=\pm \mid m \mid$ are
exactly degenerated. Therefore, each energy level with a given $l$ will be
splitted on $l+1$ closer sublevels making more dense the energy spectrum.
Furthermore, all excitations at finite temperature have associated a
finite life time. Excitations with $E \sim \mu $ which are
the ones that mainly contribute in the ``background'' transitions, have
the shortest life time.
This can be accounted for
phenomenologically by assuming that these levels have a finite
lorentzian width $\Delta $.
That is,  instead
of delta functions in the equation for the damping
rate (\ref{damping2}), we will consider a
Lorentzian distribution centered
at $\omega_{ik}$ with a fixed width $\Delta$:
$f_L(\omega_{ik},\Delta)=\Delta/(2 \pi \hbar)
[(\omega_{ik}-\Omega_{\rm osc})^2+\Delta^2/4]$.
In this case the damping rate
becomes a smooth function of $\Omega _{\rm osc}$ and its value when
$\Omega_{\rm osc}=\Omega_{\rm M}$ defines the Landau damping of the monopole
oscillations.
At conditions
\begin{equation}
\overline{\Delta \omega} \ll \Delta \ll \omega_{\rm ho}\,,
\label{Del}
\end{equation}
the damping rate will have only a weak dependence
on the exact value of $\Delta $. In Figure 4 we plot the
the dimensionless damping rate $\gamma/\Omega_{\rm M} $ as a function
of the lorentzian width $\Delta $ (in units of $\omega_{\rm ho}$)
for $N_0=50 000$ at different
temperatures.
The summation in (\ref{damping2}) has been done over all resonances
excluding of course the ``strong resonances" presented in
Figures 1, 2 and 3.
One can see that the $\Delta $-dependence is weak
indeed
in the interval $\Delta/\omega_{\rm ho}=0.05 \div 0.2 $
and $\gamma $ can be reliable extrapolated
from this interval
to the value $\Delta =0 $.
We take as Landau damping
this extrapolated value of $\gamma $.
One can estimate the accuracy of this extrapolation procedure
to be of the order of 10 \% according to the change of $\gamma $ over
this interval.
In Figure 5 we plot the damping rate versus $k_B T/\mu$ for
$N_0=150 000$ and $50 000$ atoms in the condensate.
As expected, Landau damping increases with temperature since the
number of excitations available at thermal equilibrium is larger
when $T$ increases.
One can distinguish two different regimes in Fig.~5
one at very low $T$ ($k_B T \ll \mu$) and the other at higher $T$.
The behaviour of the damping rate becomes linear at relatively small
temperature ($k_B T \sim \mu$) 
in comparison to the homogeneous system \cite{PLA} where this regime
occurs at $k_B T \gg \mu$.
Moreover, the damping rate increases for larger number of condensed
atoms because the density of states available to the system
also increases.
It is interesting to note that the order of magnitude of the damping rate
is the same as the one previously estimated for an uniform gas
\cite{Liu1,PLA,Liu2,Giorgini1}
and for anisotropic traps \cite{Fedichev1,Fedichev2}.

\section{SUMMARY}
We have considered the monopole oscillation of a Bose-condensed 
dilute atomic gas
in an isotropic trap. First of all, we have calculated the normal modes
of the condensate by solving the time-dependent Gross-Pitaevskii equation
within Bogoliubov theory \cite{PRA} and then
we have used the formalism developed in
Ref.\cite{PLA} to calculate the matrix elements associated
with the 
transitions between excited states allowed
by the monopole selection rules.
Within a first order perturbation theory we have studied the
Landau damping
of collective modes due to the coupling with thermal excited levels.
We have developed the response function formalism to study the
fluctuations of the system due to an external perturbation.
The contribution of the elementary excitations has been introduced
also perturbatively as in the calculation of the damping strength, and
we have derived analytic equations for the response function at
zero temperature and at low temperature regime.
We have seen
that when the condensate oscillates with the monopole mode and a
small perturbation to the trap frequency is applied, one can excite
new resonances at the transition frequencies.
These thermal-induced resonances are coupled with the monopole due to
interaction effects.
One cannot exclude {\it a priori} the possibility to observe
such resonances also in anisotropic traps.
This problem deserves further investigation.
Observation of these
resonances would give important and unique information about
the interaction
between elementary excitations in Bose-Einstein condensed gases.

We thank F.~Dalfovo, P.~Fedichev and S.~Stringari for helpful discussions.
M.G. thanks the Istituto Nazionale per la Fisica della Materia (Italy)
for financial support.

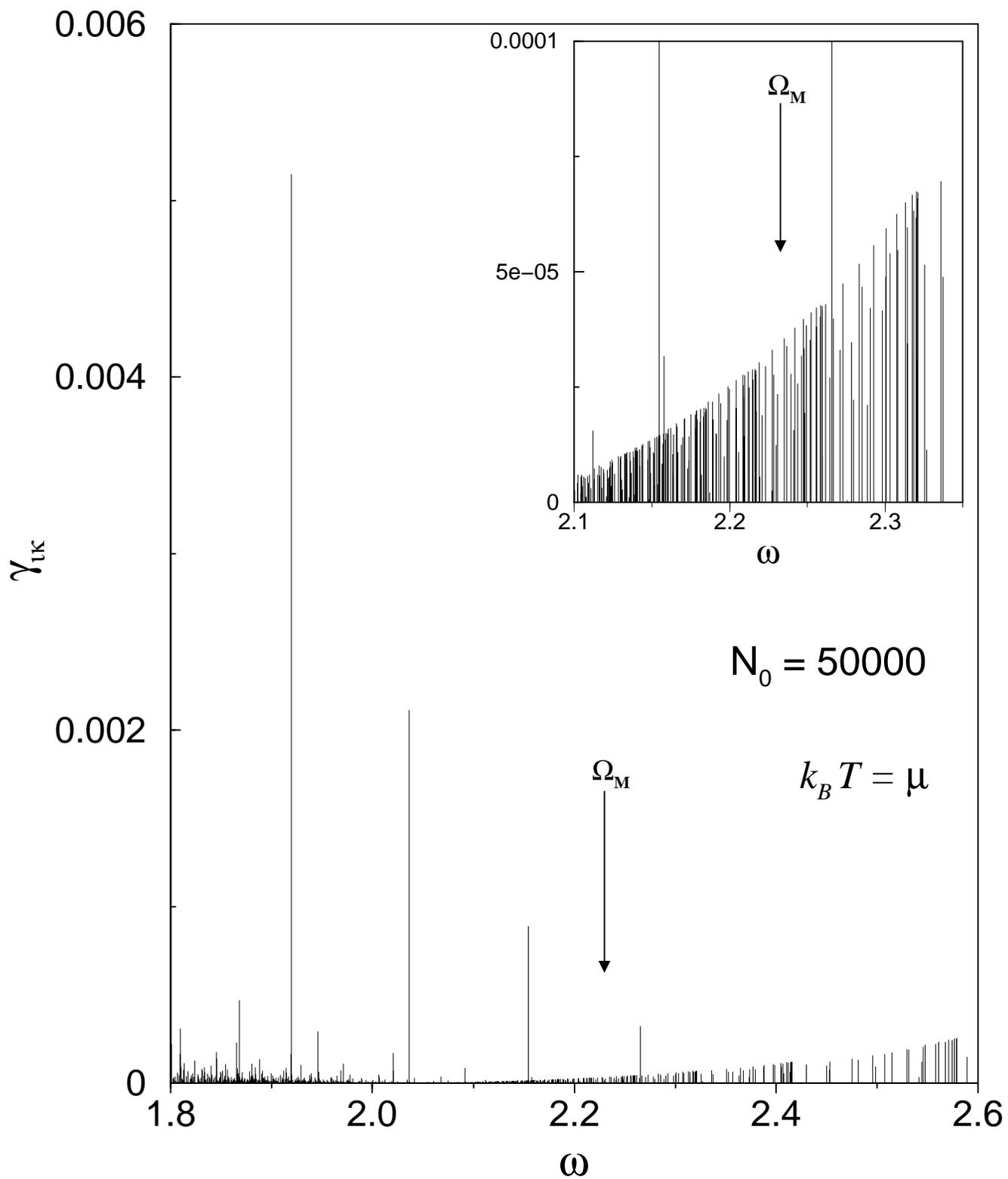
\begin{figure}
\caption{
Transition frequencies $\omega_{ik}$ (in units of $\omega_{\rm ho}$)
allowed by the monopole selection rules, for $N_0=50 000$ atoms
of $^{87}$Rb in a spherical trap with $a_{\rm ho}=0.791 \times 10^{-4}$
cm, at $k_B T = \mu$. The vertical bars have length equal to the
corresponding damping strength (in units of $\omega_{\rm ho}$).
The arrow points to the monopole frequency $\Omega_{\rm M}$.
A close-up view of the transition frequencies around $\Omega_{\rm M}$
is presented in the inset in order to show the dense background.
}
\label{fig1}
\end{figure}
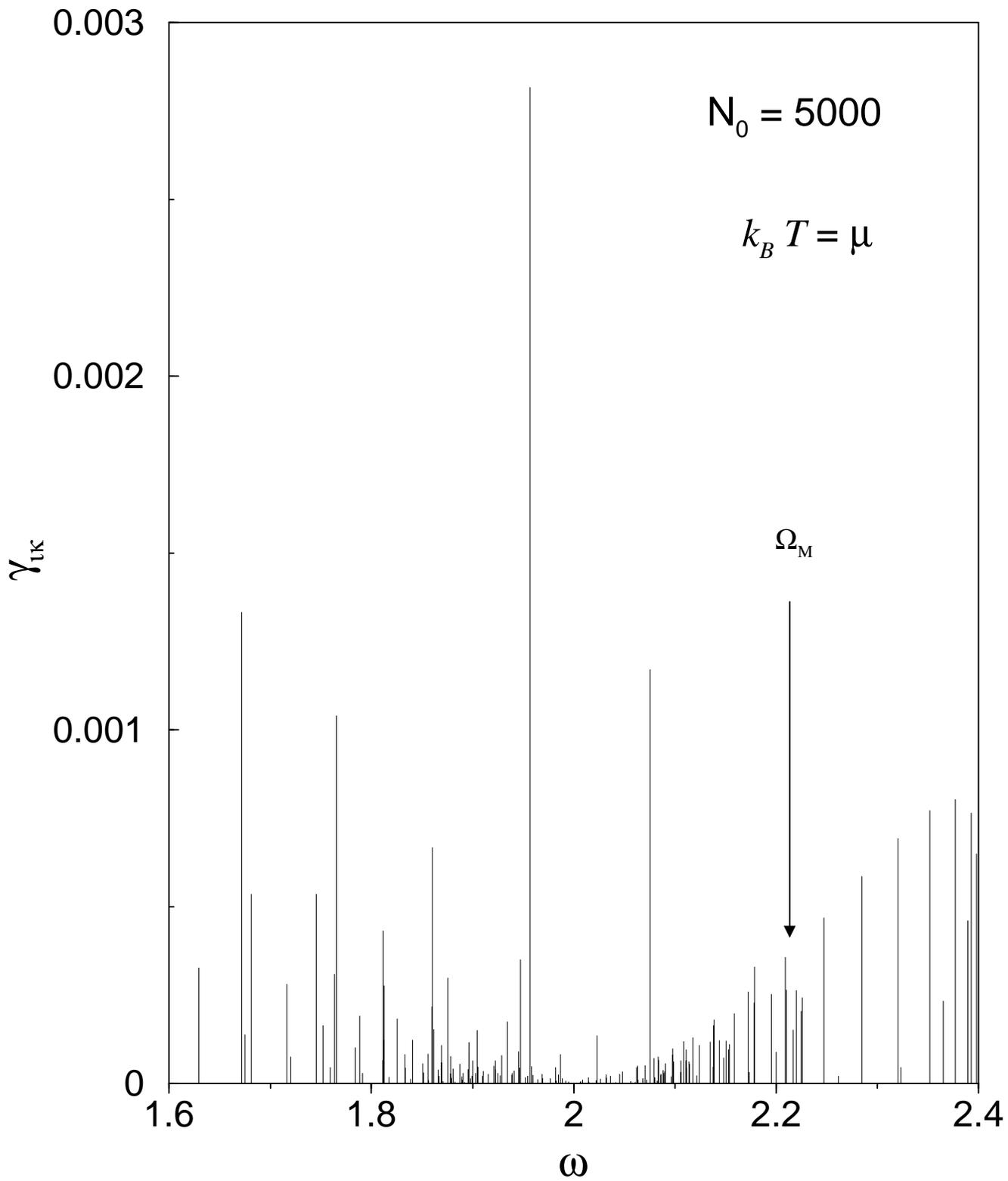
\begin{figure}
\caption{
Same as Fig. 1 but for $N_0=5 000$ atoms at $k_B T = \mu$}
\label{fig2}
\end{figure}
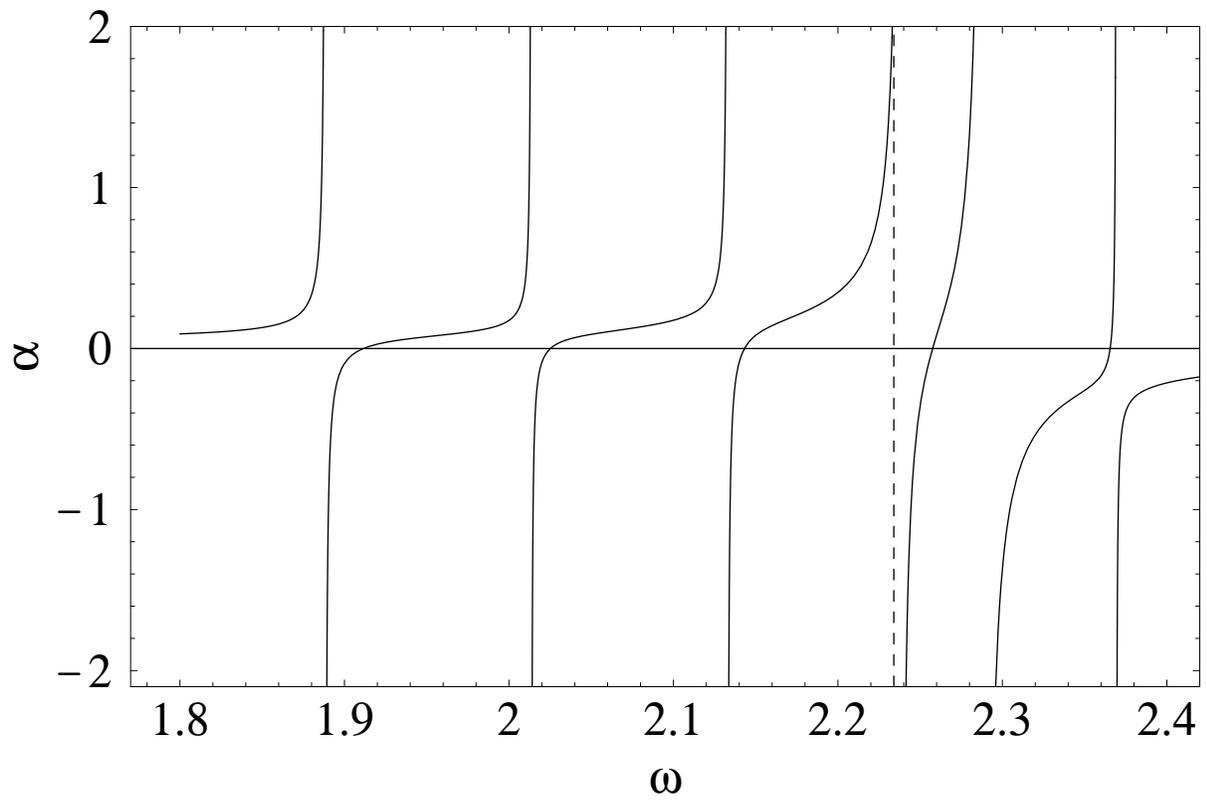
\begin{figure}
\caption{
Real part of the
response function
$\alpha (\omega)$ (in arbitrary units) as a function of $\omega$
(in units of $\omega_{\rm ho}$), for
$N_0=150 000$ at $k_B T = \mu$. The dashed line shows the
monopole resonance at $\Omega_{\rm M}$, whereas the other divergences of
$\alpha(\omega)$ correspond to the temperature-induced resonances at
$\omega_{R}$.
}
\label{fig3}
\end{figure}
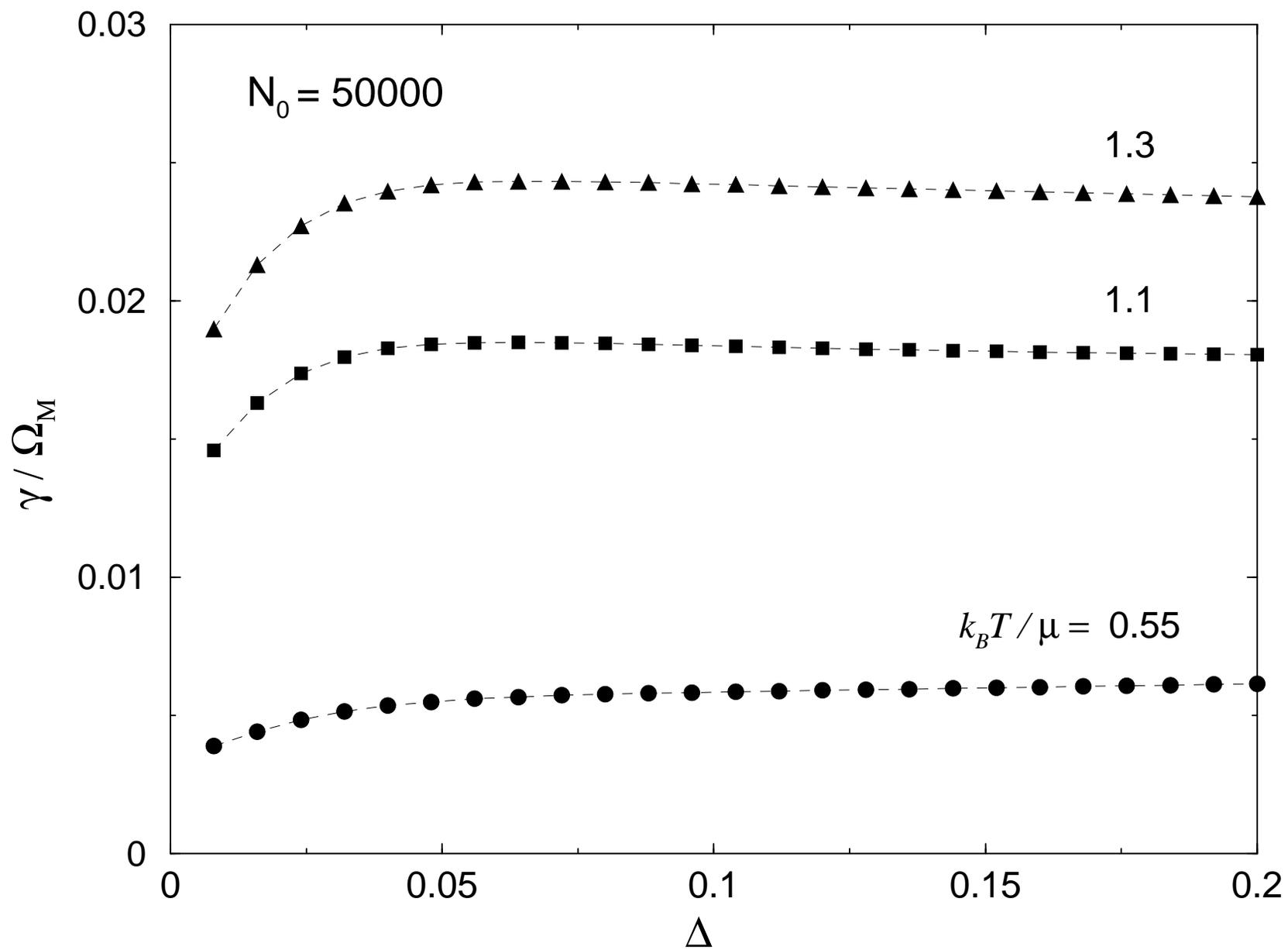
\begin{figure}
\caption{
Dimensionless damping rate $\gamma/\Omega_{\rm M}$ as a function of
the lorentzian width $\Delta$ (in units of $\omega_{\rm ho}$),
for $N_0=50 000$ $^{87}$Rb atoms in the
spherical trap at different temperatures.
Dashed lines are plotted as a guide for the eye. Solid dots,
squares and triangles, correspond to the numerical calculation
at $k_B T/\mu= 0.55$, $1.1$ and $1.3$, respectively.
}
\label{fig4}
\end{figure}
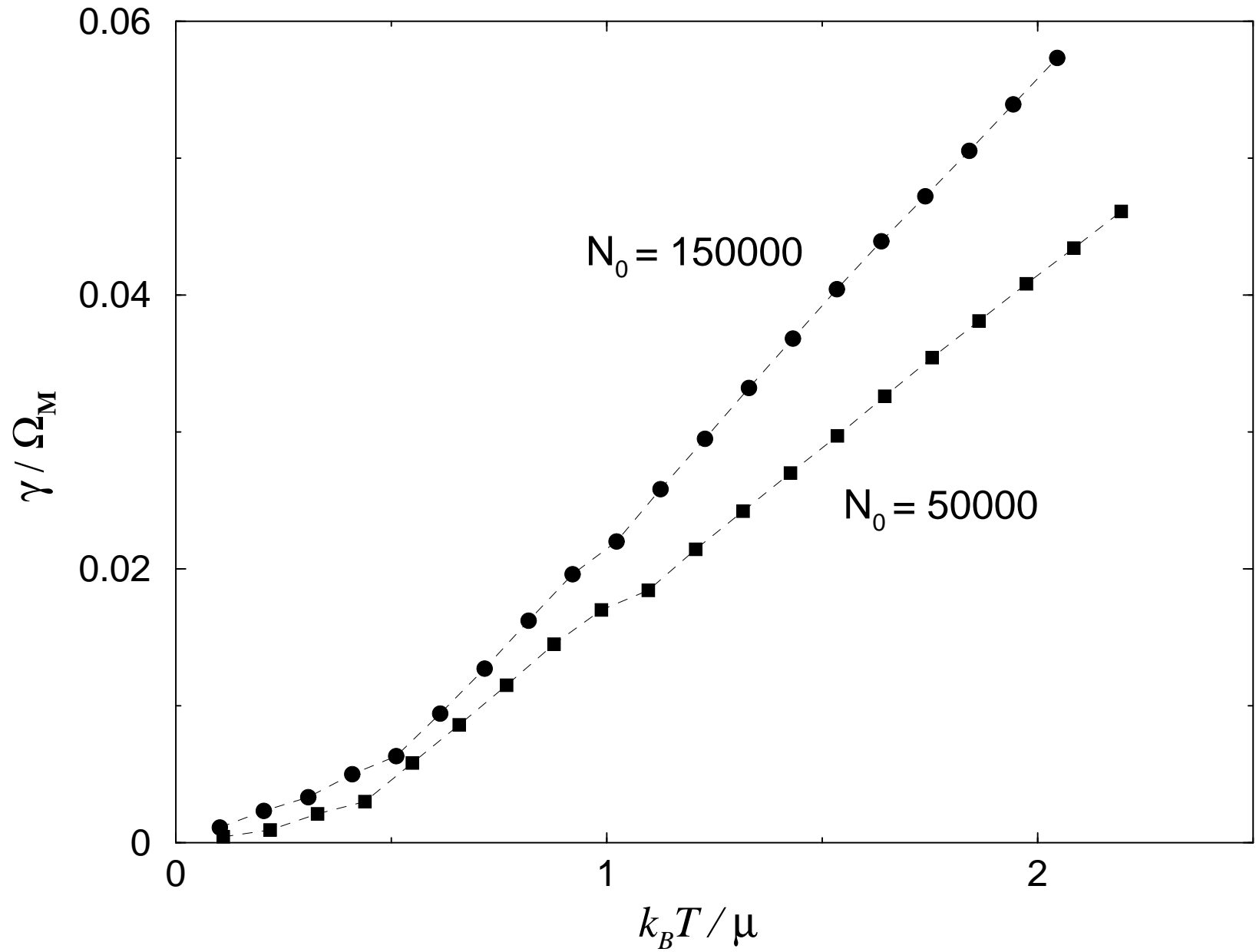
\begin{figure}
\caption{
Dimensionless damping rate $\gamma/\Omega_{\rm M}$ as a function
of $k_B T/\mu$ for
$N_0=150 000$ atoms (dots)
and $N_0=50 000$ atoms (squares).
Dashed lines are plotted as a guide for the eye.
}
\label{fig5}
\end{figure}
\begin{table}
\caption{
Damping coefficients $\gamma_R$ (in units of $\omega_{\rm ho}$) of
the ``strong resonances" $\omega_{R}$ (in units of $\omega_{\rm ho}$)
and relative intensities
$|A_{1}/A_0|$ between
temperature-induced the monopole resonance,
for $N_0=150 000$ condensate atoms of $^{87}$Rb in a spherical trap
with $a_{\rm ho}=0.791 \times 10^{-4}$ cm at $k_B T = \mu$.
}
\begin{tabular}{ccc}
$\omega_{R}$&$\gamma_{R}$&
$|A_{1}/A_0|$\\
\tableline
1.9115&0.009268&0.063\\
2.0252&0.004147&0.063\\
2.1432&0.002102&0.127\\
2.2576&0.001097&0.298\\
2.3655&0.000545&0.020\\
\end{tabular}
\label{table1}
\end{table}


\begin{references}

\bibitem{JILA}D.S. Jin, M.R. Matthews, J.R. Ensher, C.E. Wieman and
E.A. Cornell, Phys. Rev. Lett. {\bf 78} (1997) 764.

\bibitem{MIT} M.-O. Mewes, M.R. Anderson, N.J. van Druten, D.M. Kurn, D.S.
Durfee, C.G. Townsend and W. Ketterle, Phys. Rev. Lett. {\bf 77} (1996)
988.

\bibitem{Liu1}W.V. Liu and W.C. Schieve, cond-mat/9702122 preprint.

\bibitem{PLA}L.P. Pitaevskii and S. Stringari, Phys. Lett. A {\bf 235}
(1997)
398.

\bibitem{Liu2}W.V. Liu, Phys. Rev. Lett. {\bf 79} (1997) 4056.

\bibitem{Giorgini1}S. Giorgini, Phys. Rev. A {\bf 57} (1998) 2949.

\bibitem{Fedichev1}P.O. Fedichev, G.V. Slyapnikov, J.T.M. Walraven,
Phys. Rev. Lett. {\bf 80} (1998) 2269.

\bibitem{Fedichev2}P.O. Fedichev, G.V. Slyapnikov, Phys. Rev. A {\bf 58}
(1998) 3146.

\bibitem{bog} N.N.~Bogoliubov, J. Phys. USSR, {\bf 11}, 23 (1947).

\bibitem{G}  E.P.~Gross, Nuovo Cimento
{\bf 20}, 454 (1961); E.P.~Gross, J. Math. Phys. {\bf 4}, 195 (1963)

\bibitem{P} L.P.~Pitaevskii, Zh. Eksp. Teor. Fiz. {\bf 40}, 646 (1961)
[Sov. Phys. JETP {\bf 13}, 451 (1961)].

\bibitem{KetVa} W. Ketterle, D.S. Durfee, and D.M. Stamper-Kurn, in
{\it Proceedings of Int. School E. Fermi}, Varenna 1998,
cond-mat/9904034.

\bibitem{Edwards} M. Edwards, P.\ A.\ Ruprecht,
K.\ Burnett, R.\ J.\ Dodd, and C.\ W.\ Clark,
Phys. Rev. Lett. {\bf 77}, 1671 (1996);  P. A. Ruprecht,
Mark Edwards, K. Burnett, and Charles W. Clark,
Phys. Rev. A {\bf 54}, 4178 (1996) ;  M. Edwards,
R.\ J.\ Dodd, C.\ W.\ Clark, and K. Burnett, J. Res. Natl.
Inst. Stand. Technol. {\bf 101}, 553 (1996).

\bibitem{Singh} K.G. Singh and D.S. Rokhsar, Phys. Rev. Lett.
{\bf 77}, 1667 (1996)

\bibitem{Esry} B.D. Esry, Phys. Rev. A {\bf 55}, 1147 (1997)

\bibitem{Zaremba} D.A.W. Hutchinson, E. Zaremba, and A. Griffin,
Phys. Rev. Lett. {\bf 78}, 1842 (1997)

\bibitem{You}  L. You, W. Hoston, and M. Lewenstein,
Phys. Rev. A {\bf 55}, R1581 (1997)

\bibitem{PRA}F. Dalfovo, S. Giorgini, M. Guilleumas, L.P. Pitaevskii and
S. Stringari, Phys. Rev. A {\bf 56} (1997) 3840.

\bibitem{Stringari} S. Stringari, Phys. Rev. Lett. {\bf 77} (1996) 2360.

\bibitem{hydro} When $Na/a_{\rm ho}$ is large one can use the
hydrodynamic approximation for the functions $u$ and $v$
of the low-energy excitations [see
L.P. Pitaevskii, {\it Recent Progress in Many-Body Theories},
Ed. D. Nielson and R. Bishop, World Scientific (Singapore, 1998), p.3].
However, we have found that the matrix elements $A_{ik}$, defined in
section III, are quite sensitive to the accuracy of these functions.
Hydrodynamics can give values for
$\gamma_{ik}$ an order of magnitude smaller than the values calculated
numerically by using the Bogoliubov functions
(\ref{coupled1},\ref{coupled2}).
On the contrary, it is
completely safe to use hydrodynamics to calculate the response function
at zero temperature $\alpha _0(\omega)$
(see section IV),
since the difference between hydrodynamics and exact values
of the monopole frequency $\Omega_{\rm M}$ is very small.

\bibitem{Beliaev} Beliaev decay
of an elementary excitation into a pair of excitations
[S.T. Beliaev, Soviet Phys. JETP {\bf 34} (1958) 323]
is not active for the lowest energy modes in the case of trapping
potential because of the discretization of levels.

\bibitem{Landau}L.D. Landau and E.M. Lifshitz, {\it Statistical Physics,
Course of theoretical Physics} (Vol. 5), Pergamon Press (1970).

\bibitem{Lev} L.P. Pitaevskii, Phys. Lett. A {\bf 229} (1997) 406.

\bibitem{Concetta1} F. Dalfovo, C. Minniti and L.P. Pitaevskii,
Phys. Rev. A {\bf 56} (1997) 4855.

\bibitem{Dalfovo} F. Dalfovo, S. Giorgini, L.P. Pitaevskii and S.
Stringari, Rev. Mod. Phys. {\bf 71} (1999) 463.


\bibitem{Abrikosov} A.A. Abrikosov, L.P. Gorkov and I. Ye.
Dzyaloshinskii, {\it Quantum Field Theoretical Methods in Statistical
Physics} (Pergamon Press, 1965)

\bibitem{Giorgini3} S. Giorgini, L.P. Pitaevskii and S. Stringari,
J. Low Temp. Phys. {\bf 109} (1997) 309.

\bibitem{Popov} In this
paper we do not include the so-called Popov's self-consistent
correction in equation (\ref{TDGP}). See, for
example, Ref.~\cite{Giorgini3}.

\bibitem{dipole}
We have neglected all the transitions
that involve the lowest dipole mode ($l=1$, $n=0$) because in an
external harmonic potential this mode, corresponding to the oscillation
of the center of mass, is unaffected
by the interatomic forces and then the transition probability due to
interaction effects (\ref{prob}) must be zero \cite{Stringari,Stoof}.
However, the perturbation formalism
we have presented in section III does not take into account this
physical consideration automatically and therefore, we have omitted
by hand all transitions with the dipole mode.

\bibitem{Stoof} H.~Stoof, J. Low Temp. Phys. , {\bf 114},
11 (1999).






\end{references}
\end{document}